\documentclass[american,aps,pra,superscriptaddress,twocolumn]{revtex4-2}

\usepackage[T1]{fontenc}
\usepackage[sc]{mathpazo}
\usepackage{amsmath}
\usepackage{amssymb}
\usepackage{enumerate}
\usepackage{amsthm}
\usepackage{color}
\usepackage{graphicx} 

\usepackage{hyperref}
\hypersetup{pdfpagemode=UseNone}

\newcommand{\ket}[1]{|#1\rangle}

\newcommand{\braket}[1]{\langle#1\rangle}

\begin{document}

\title{The adiabatic theorem for non-Hermitian quantum systems with real eigenvalues and
the complex geometric phase}

 \author{Minyi Huang}
 \email{11335001@zju.edu.cn}
 \email{hmyzd2011@126.com}
 \affiliation{Department of Mathematical Sciences, Zhejiang Sci-Tech University, Hangzhou 310018, PR~China}

\author{Ray-Kuang Lee}
\email{rklee@ee.nthu.edu.tw}
\affiliation{Department of Physics, National Tsing Hua University, Hsinchu 300, Taiwan}
\affiliation{Institute of Photonics Technologies, National Tsing Hua University, Hsinchu 300, Taiwan}

\begin{abstract}
 The adiabatic theorem is one of the most interesting and significant theorems in quantum mechanics. However, the adiabatic theorem can fail for general non-Hermitian quantum systems.
 In this paper, by utilizing the complex geometric phase, the functional calculus for biorthogonal systems and the Gr\"{o}nwall inequality, we prove rigorously that the adiabatic theorem is still valid for diagonalizable non-Hermitian systems with real eigenvalues. The proof also justifies the definition of a complex Berry phase for non-Hermitian systems, in both Abelian and non-Abelian cases.
\end{abstract}

\maketitle

\section{Introduction}
\label{sec:intro}


 An interesting and important issue of time-dependent quantum systems is the adiabatic theorem.
 The theorem states that if a quantum system evolves slowly enough under a time-dependent non-degenerate
 Hamiltonian and the initial state is in the instantaneous eigenstate of the Hamiltonian, then the final state of
 the system will remain in the instantaneous eigenstate up to a multiplicative phase factor \cite{born1928beweis}.
 In 1950, T. Kato gave an elegant but technically involved proof of the theorem \cite{kato1950adiabatic}.
Since its discovery, the adiabatic theorem has attracted much interests from researchers, and many related topics were discussed \cite{avron1999adiabatic,teufel2001note,tong2010quantitative}. In particular, it is intimately related to the studies of geometric phase and quantum computing, with a wide range of applications~\cite{berry1984quantal,farhi2001quantum,albash2018adiabatic}.
However, it should be mentioned that the original discussions on the adiabatic theorem does not involve the Berry phase, which
is now usually considered as an indispensable companion with the adiabatic evolution. In fact, it is more than fifty years later when Berry systematically investigated the geometric phase bearing his name~\cite{berry1984quantal}. Later, the concept of  geometric phase is also
extended to the non-abelian case \cite{wilczek1984appearance}.

The discussions of non-Hermitian quantum systems has a long history and recently there is a growing interests in such systems,
for their novel features and potential applications \cite{bender1998real,bender2007making,mostafazadeh2010pseudo}.
In particular, the dynamics and topology of non-Hermitian systems, such as the skin effect, has attracted much attentions~\cite{ashida2020non,PhysRevLett.121.086803,PhysRevLett.121.136802,alvarez2018non,ozdemir2019parity,PhysRevLett.123.170401,PhysRevLett.123.246801,PhysRevResearch.1.023013}.
The original discussions of the adiabatic theorem is within the framework of Hermitian systems.
Recently there are discussions on the adiabatic theorem and geometric phase in
non-Hermitian systems \cite{garrison1988complex,nenciu1992adiabatic,mehri2008geometric,berry2011slow,jing2016eigenstate,
gong2019piecewise,jin2025universal,kumar2025general}.
 A key difference between the adiabatic evolution of Hermitian and non-Hermitian systems is that the adiabatic theorem may
fail for the latter ones.

In this paper, we discuss the adiabatic following problem, with the focus on diagonalizable Hamiltonians with real
eigenvalues.
By utilizing a complex geometric phase, the Gr\"{o}nwall inequality and generalizing Kato's functional calculus techniques to
biorthogonal systems, we prove rigorously that the adiabatic theorem is still valid for non-Hermitian systems with
 real eigenvalues, in both Abelian and non-Abelian cases.
The structure of the paper is organized as follows. In section II, we briefly review the adiabatic theorem.
In section III, it is shown that the adiabatic theorem is always valid for diagonalizable Hamiltonians with real
eigenvalues. It also shows that the geometric phase is generally dependent on the specific choice of basis, but gives the
same evolution of an eigenstate in its all different representations.
In section IV, we give some discussions, show the insights for the result, and conclude this paper.
Section V is the appendix.

\section{Brief review of adiabatic theorem}
Usually, the adiabatic following problem is formulated in a Hermitian system. We briefly review the discussions of the theorem, following the way in \cite{kato1950adiabatic,messiah2014quantum}.\\

Let $H(s) (0\leq s\leq 1)$ be a collection of Hermitian Hamiltonians. Now by setting $s=\frac{t}{T}$, where $t$ is a new
variable and $T$ is a parameter. Apparently, $H(\frac{t}{T})$ changes slower than $H(s)$ if $T\geq 1$. Consider the
Schr\"{o}dinger equation of $H(\frac{t}{T})$,
\[
i \frac{d}{dt} U_T(t)=H(\frac{t}{T})U_T(t),
\]
where $U_T(t)$ is the abbreviation of $U_T(t,0)$, the evolution operator of $H(\frac{t}{T})$. The subscript letter $T$ implies the fact that $U_T(t)$ is generally dependent of the parameter $T$. By rewriting the Schr\"{o}dinger equation using the variable $s=\frac{t}{T}$, one can obtain
\begin{equation}
i \frac{d}{ds} U_T(s)=T H(s)U_T(s).\label{eqn:adiabatic}
\end{equation}
Note that when deriving Eq. (\ref{eqn:adiabatic}), one does not bother to make a difference in the notation between $U_T(t)$ and $U_T(s)$,
which usually does not cause confusion.
The adiabatic theorem discuss the behaviour of $U_T(t)$, or equivalently $U_T(s)$, when $T$ tends to infinity. A mathematical formalism of the theorem is
\begin{equation}
\lim_{T\rightarrow\infty} U_T(s)\ket{j(0)}=P_j(s)\lim_{T\rightarrow\infty}U_T(s)\ket{j(0)},\label{eqn:adiabatic m}
\end{equation}
where $\ket{j(s)}$ is the $j$-th eigenstate of $H(s)$ and $P_j(s)$ is the corresponding orthogonal projection.
Briefly speaking, when the initial state is set in $\ket{j(0)}$ and evolves slowly enough, the theorem or equivalently Eq. (\ref{eqn:adiabatic m}) guarantees that the state follows the $j$-th eigenstate $\ket{j}$ up to a factor. On the other hand,
in the degenerate case, the state will remain in the subspace spanned by the eigenvectors of the same eigenvalue, which introduced the non-abelian geometric phase \cite{wilczek1984appearance}.
The proof of the adiabatic theorem utilizes the properties of orthogonal projections and unitary operators can be found in \cite{kato1950adiabatic,messiah2014quantum}. Thus when discussing the non-Hermitian quantum systems, the proof is not directly applicable.

\section{The main result}
\subsection{The generalized adiabatic theorem and its proof}
In this section, we prove the generalized adiabatic theorem. Like previous literatures in this field, it is always assumed that
the functions in the following discussion satisfy the necessary condition of continuity \cite{kato1950adiabatic,messiah2014quantum}. \\~\\
\emph{The adiabatic theorem is valid for diagonalizable Hamiltonians with real eigenvalues.}\\

Let $H(s)$ be a Hamiltonian satisfying
 \begin{equation}
H(s)v_i(s)=\lambda_i(s)v_i(s),
\end{equation}
where $\lambda_i(s)$ and $v_i(s)$ are the real eigenvalues and corresponding eigenvectors, respectively. Now let $\xi_i(s)$ be the
corresponding biorthogonal vectors, i.e. $\xi_i^\dag(s)v_i(s)=\delta_{ij}.$ Note that $\xi_i(s)$ always exist. In fact, denote
$V(s)$ the matrix whose column vectors are $v_i(s)$, then $\xi_i^\dag(s)$ are the row vectors of the inverse of $V(s)$.
Since the eigenvalues can be degenerate, $H(s)$ and $I$ can be rewritten as
 \begin{eqnarray}
&&H(s)=\sum_k\lambda_k(s)P_k(s),\label{resol1}\\
&&I=\sum_k P_k(s),\label{resol2}
\end{eqnarray}
 where $\lambda_k(s)$ are the different eigenvalues and $P_k(s)$ are the projections such that
$P_k(s)P_l(s)=\delta_{kl}P_k(s)$. To see the existence of $P_k(s)$, note that one can take the eigenvectors of the same eigenvalue together. For example, let the eigenvectors of $\lambda_k(s)$ be $v_{j_k}(s)$, then $P_k(s)=\sum_{j_k} v_{j_k}(s)\xi^\dag_{j_k}(s)$.
Moreover, it can verified that $P_k(s)$ is independent of the choice of $v_{j_k}(s)$. In fact, if one takes $u_{j_k}(s)$ and $\zeta_{j_k}(s)$, then direct calculations show that $P_k(s)=\sum_{j_k} v_{j_k}(s)\xi^\dag_{j_k}(s)=\sum_{j_k} u_{j_k}(s)\zeta^\dag_{j_k}(s)$ since $v_{j_k}(s)$, $u_{j_k}(s)$ are related by an invertible matrix and
$\xi_{j_k}(s)$, $\zeta_{j_k}(s)$ are related by its inverse matrix.

Now for any $j$, one can define an operator $S(s)$ by
 \begin{equation}
S(s)=\sum_{k\neq j}\frac{1}{\lambda_k(s)-\lambda_j(s)}P_k(s),\label{S}
\end{equation}
which satisfies that following relation
\begin{eqnarray}
&&S(s)P_j(s)=P_j(s)S(s)=0,\label{s1}\\
&&[H(s)-\lambda_j(s)I]S(s)=I-P_j(s).\label{s2}
\end{eqnarray}

Now suppose that the evolution operator is $U_T(s)$, then by Eq. (\ref{eqn:adiabatic})
\begin{equation}
i\frac{d}{ds}U_T^{-1}(s)=-U_T^{-1}(s)TH(s).
\label{eqn:adiabatic i}
\end{equation}
One can further obtain
\begin{eqnarray}
\nonumber&&\frac{d}{ds}[e^{-iT\int_0^s\lambda_j(\sigma)d\sigma}U_T^{-1}(s)]\\
\nonumber&=&\frac{d}{ds}[e^{-iT\int_0^s\lambda_j(\sigma)d\sigma}]U_T^{-1}(s)+e^{-iT\int_0^s\lambda_j(\sigma)d\sigma}\frac{d}{ds}[U_T^{-1}(s)]\\
\nonumber&=&-iT\lambda_j(s)e^{-iT\int_0^s\lambda_j(\sigma)d\sigma}U_T^{-1}(s)\\
\nonumber&+&e^{-iT\int_0^s\lambda_j(\sigma)d\sigma}U_T^{-1}(s)iTH(s)\\
&=&iTe^{-iT\int_0^s\lambda_j(\sigma)d\sigma}U_T^{-1}(s)[H(s)-\lambda_j(s)I].\label{derivative3}
\end{eqnarray}
On the other hand, for any eigenvector $v_{j_l}(s)$ of $\lambda_j(s)$, one can prove that there exists some operator $K_j(s)$ satisfying
\begin{eqnarray}
\frac{d}{ds}[K_j(s)v_{j_l}(s)]=[I-P_j(s)]\frac{d}{ds}[K_j(s)v_{j_l}(s)].
\label{derivative}
\end{eqnarray}
In fact, Eq. (\ref{derivative}) is equivalent to
\begin{eqnarray}
P_j(s)\frac{d}{ds}[K_j(s)v_{j_l}(s)]=0.
\label{derivativeP}
\end{eqnarray}
To uniquely determine $K_j(s)$, we assume that $\lambda_j$ is an eigenvalue of multiplicity $h$ and $K_j(s)v_{j_l}(s)$ is always in the subspace spanned by the eigenvectors of $\lambda_j$. In this case,
$K_j(s)$ can be rewritten as $K_j(s)=\sum_{mn}K^{mn}_{j}(s)v_m(s)\xi_n^\dag(s)$, where $m,n =j_1,j_2...,j_h$. Then Eq. (\ref{derivativeP}) can be rewritten as
\begin{eqnarray}
 \frac{d}{ds}[\tilde{K}_j(s)]+G(s)\tilde{K}_j(s)=0,\label{derivativeP2}
\end{eqnarray}
where $\tilde{K}_j(s)$ is the matrix consisting of coefficients $K^{mn}_j(s)$,
$G(s)$ is the matrix consisting of coefficients $G_{mn}(s)=\xi^\dag_{m}(s)\dot{v}_{n}(s)$. Note that $G(s)$ is actually the
(non-Abelian) geometric phase, reducing to the classical Berry phase in the non-degenerate case.

With the initial value condition $\tilde{K}_j(0)=I$ and the continuity of the eigenvectors,
Eq.(\ref{derivativeP2}) gives a unique solution of $\tilde{K}_j(s)$, and thus $K_j(s)$ to Eq. (\ref{derivative}). In fact, such a solution is often represented by the time ordering operator \cite{wilczek1984appearance}. On the other hand, it should be mentioned that the uniqueness of $K_j(s)$ is only valid when a specific choice of basis is fixed. That is, different basis will generally give different operator $K_j(s)$. However, as we shall see, $K_j(s)v_{j_l}(s)$ is independent of the specific choice of basis $v_{j_l}(s)$ up to some constant factor, leading to the same evolution effect with respect to different basis (see Section III B for details). This also justifies the definition of geometric phase.

Now by substituting Eq. (\ref{s2}) into Eq. (\ref{derivative}), we have
\begin{eqnarray}
\nonumber&&\frac{d}{ds}[K_j(s)v_{j_l}(s)]\\
\nonumber&=&[I-P_j(s)]\frac{d}{ds}[K_j(s)v_{j_l}(s)]\\
\nonumber&=&[H(s)-\lambda_j(s)I]S(s)\frac{d}{ds}[K_j(s)v_{j_l}(s)].
\label{derivative2}\\
\end{eqnarray}

For simplicity, denote $K_j(s)v_{j_l}(s)$ by $\tilde{v}_{j_l}(s)$.
Eq. (\ref{derivative2}) can be rewritten as
\begin{eqnarray}
\frac{d}{ds}[\tilde{v}_{j_l}(s)]=[H(s)-\lambda_j(s)I]S(s)\frac{d}{ds}[\tilde{v}_{j_l}(s)].
\label{derivative2n}
\end{eqnarray}
Then we have
\begin{eqnarray}
\nonumber&&\frac{d}{ds}[e^{-iT\int_0^s\lambda_j(\sigma)d\sigma}U_T^{-1}(s)\tilde{v}_{j_l}(s)]\\
\nonumber&=& \frac{d}{ds}[e^{-iT\int_0^s\lambda_j(\sigma)d\sigma}U_T^{-1}(s)]\tilde{v}_{j_l}(s)\\
&+& e^{-iT\int_0^s\lambda_j(\sigma)d\sigma}U_T^{-1}(s)\frac{d}{ds}[\tilde{v}_{j_l}(s)]\label{derivative1}
\end{eqnarray}
By Eqs. (\ref{derivative3}) and (\ref{s1}), the first term in Eq. (\ref{derivative1}) is
\begin{eqnarray}
\nonumber&& \frac{d}{ds}[e^{-iT\int_0^\sigma\lambda_j(\sigma)d\sigma}U_T^{-1}(s)]\tilde{v}_{j_l}(s)\\
\nonumber&=&iTe^{-iT\int_0^s\lambda_j(\sigma)d\sigma}U_T^{-1}(s)[H(s)-\lambda_j(s)I]\tilde{v}_{j_l}(s)\\
&=&0.
\end{eqnarray}
Thus Eq. (\ref{derivative1}) reduces to
\begin{eqnarray}
\nonumber&&\frac{d}{ds}[e^{-iT\int_0^s\lambda_j(\sigma)d\sigma}U_T^{-1}(s)\tilde{v}_{j_l}(s)]\\
&=& e^{-iT\int_0^s\lambda_j(\sigma)d\sigma}U_T^{-1}(s)\frac{d}{ds}[\tilde{v}_{j_l}(s)].
\label{zero}
\end{eqnarray}
Now by taking integral in Eq. (\ref{zero}),
\begin{eqnarray}
\nonumber&&e^{-iT\int_0^s\lambda_j(\sigma)d\sigma}U_T^{-1}(s)\tilde{v}_{j_l}(s)-\tilde{v}_j(0)\\
\nonumber&=&\int_0^s \frac{d}{d\sigma}[e^{-iT\int_0^\sigma\lambda_j(r)dr}U_T^{-1}(\sigma)\tilde{v}_{j_l}(\sigma)]d\sigma\\
\nonumber&=&\int_0^s e^{-iT\int_0^\sigma\lambda_j(r)dr}U_T^{-1}(\sigma)\frac{d}{d\sigma}[\tilde{v}_{j_l}(\sigma)]d\sigma.
\end{eqnarray}
By Eqs. (\ref{derivative2n}), (\ref{derivative3}) and integration by parts, it follows that
\begin{eqnarray}
\nonumber&&e^{-iT\int_0^s\lambda_j(\sigma)d\sigma}U_T^{-1}(s)\tilde{v}_{j_l}(s)-\tilde{v}_j(0)\\
\nonumber&=&\int_0^s e^{-iT\int_0^\sigma\lambda_j(r)dr}U_T^{-1}(\sigma)\frac{d}{d\sigma}[\tilde{v}_{j_l}(\sigma)]d\sigma\\
\nonumber&=&\int_0^s e^{-iT\int_0^s\lambda_j(r)dr}U_T^{-1}(\sigma)(H(\sigma)-\lambda_j(\sigma)I)S(\sigma)\frac{d}{d\sigma}[\tilde{v}_{j_l}(\sigma)]d\sigma\\
\nonumber&=&\frac{1}{iT}\int_0^s \frac{d}{d\sigma}[e^{-iT\int_0^\sigma\lambda_j(r)dr}U_T^{-1}(\sigma)]S(\sigma)\frac{d}{d\sigma}[\tilde{v}_{j_l}(\sigma)]d\sigma\\
\nonumber&=&\frac{1}{iT}[e^{-iT\int_0^\sigma\lambda_j(r)dr}U_T^{-1}(\sigma)]S(\sigma)\frac{d}{d\sigma}[\tilde{v}_{j_l}(\sigma)]|_0^s\\
\nonumber&-&\frac{1}{iT}\int_0^s e^{-iT\int_0^\sigma\lambda_j(r)dr}U_T^{-1}(\sigma)\frac{d}{d\sigma}[S(\sigma)\frac{d}{d\sigma}[\tilde{v}_{j_l}(\sigma)]]d\sigma.\\
\label{adiabatic1}
\end{eqnarray}
As we shall prove later, $U_T^{-1}(s)$ is uniformly bounded with respect to $T$. Then the Eq. (\ref{adiabatic1}) tends to vanish and one can see that
\begin{eqnarray}
e^{-iT\int_0^s\lambda_j(\sigma)d\sigma}U_T^{-1}(s)\tilde{v}_{j_l}(s)-\tilde{v}_{j_l}(0)\rightarrow 0.
\label{adiabatic0}
\end{eqnarray}
Moreover, as we shall prove later, $U_T(s)$ is also uniformly bounded with respect to $T$.
Thus by multiplying $U_T(s)$ on both sides of (\ref{adiabatic0}), one can obtain
 \begin{eqnarray}
 U_T(s)v_{j_l}(0)\rightarrow e^{-iT\int_0^s\lambda_j(\sigma)d\sigma}K_j(s)v_{j_l}(s).
\label{Utvjg}
\end{eqnarray}
This is exactly the adiabatic theorem. When the initial state is the eigenstate and the system evolves slowly enough, the final
state remains in the subspace of instantaneous eigenstates. In particular, when the eigenvalues are non-degenerate, one has
 \begin{eqnarray}
 \nonumber U_T(s)v_j(0)\rightarrow e^{-iT\int_0^s\lambda_j(\sigma)d\sigma}e^{-\int_0^s\braket{\xi_j(\sigma)|\dot{v}_j(\sigma)}dr}v_j(s),\\
\label{Utvj2}
\end{eqnarray}
the state remains an eigenstate up to a factor. The factor consists of two parts, one is the dynamic phase and the
other is the Berry phase.
\subsection{The evolution is independent of the choice of basis}
We first derive Eq. (\ref{derivativeP2}) from Eq. (\ref{derivativeP}), and then show Eq. (\ref{derivativeP2}) gives a unique operator $K_j(s)$ for the specific basis.

As mentioned in section III A, we assume that $K_j(s)=\sum_{mn}K^{mn}_{j}(s)v_m(s)\xi_n^\dag(s)$, where $m,n =j_1,j_2...,j_h$.
 Moreover, $P_j(s)=\sum v_i(s)\xi_i^\dag(s)$, where $i=j_1,j_2...j_h$.
 Now by expanding Eq. (\ref{derivativeP}), we have
\begin{eqnarray}
&&P_j(s)\frac{d}{ds}[K_j(s)v_{j_l}(s)]\nonumber\\
&=&\sum_i v_i(s)\xi_i^\dag(s)\frac{d}{ds}[\sum_{mn}K^{mn}_{j}(s)v_m(s)\xi_n^\dag(s)v_{j_l}(s)]\nonumber\\
&=&\sum_i v_i(s)\xi_i^\dag(s)\frac{d}{ds}[\sum_{m}K^{mj_l}_{j}(s)v_m(s)]\nonumber\\
&=&\sum_{m}\frac{d}{ds}[K^{mj_l}_{j}(s)]v_m(s)+\sum_{im}K^{mj_l}_{j}(s)\xi_i^\dag(s)v_m'(s)v_i(s).\nonumber\\
\label{derivativeP22}
\end{eqnarray}
For each $j_k=j_1,j_2...j_h$, multiply $\xi_{j_k}^\dag(s)$ on both sides of Eq. (\ref{derivativeP22}), then
\begin{eqnarray}
\frac{d}{ds}[K^{j_kj_l}_{j}(s)]+\sum_{m}K^{mj_l}_{j}(s)\xi_{j_k}^\dag(s)v_m'(s).\label{derivativeP3}
\end{eqnarray}
Note that Eq. (\ref{derivativeP3}) is valid for any $j_k, j_l=j_1,j_2...j_h$. Now by the multiplication rule of matrices, we see immediately that
Eq. (\ref{derivativeP3}) is equivalent to Eq. (\ref{derivativeP2}).\\

To see that the operator $K_j(s)$ depends on the choice of basis but gives a unique evolution, assume that $u_i(s)$ is another basis of the subspace and $\zeta_i^\dag(s)u_i(s)=\delta_{ij}$. Moreover, $u_i(s)=\sum_j X_{ji}(s)v_j(s)$, where the coefficients $X_{ji}(s)$ give rise to an invertible matrix $X(s)$. A direct calculation shows that $\zeta_i^\dag(s)=\sum_j X_{ij}^{-1}(s)\xi_j^\dag(s)$.
Now according to Eq. (\ref{derivativeP2}), the two basis give two differential equations of matrices respectively,
\begin{eqnarray}
\frac{d}{ds}[\tilde{K}_{j}^{(1)}(s)]+G^{(1)}(s)\tilde{K}_{j}^{(1)}(s)=0,\label{G1}\\
\frac{d}{ds}[\tilde{K}_{j}^{(2)}(s)]+G^{(2)}(s)\tilde{K}_{j}^{(2)}(s)=0,\label{G2}
\end{eqnarray}
where $G^{(1)}_{mn}(s)=\xi^\dag_{m}(s)\dot{v}_{n}(s)$, $G^{(2)}_{mn}(s)=\zeta^\dag_{m}(s)\dot{u}_{n}(s)$, $m,n=j_1,j_2...j_h$.
By the above equations, we see that Eq. (\ref{G2}) reduces to
\begin{eqnarray}
K_2'+X^{-1}G^{(1)}XK_2+X^{(-1)}X'K_2=0,\label{K12X}
\end{eqnarray}
where we omit the variable $s$ and denote $\tilde{K}_{j}^{(i)}(s)$ by $K_i$ for simplicity.
One can further show that
\begin{equation}
\frac{d}{ds}[K_1^{-1}(s)X(s)K_2(s)]=0.\label{d1}
\end{equation}
 Indeed, according to Eq. (\ref{G1}),
\begin{eqnarray}
&&\frac{d}{ds}[K_1^{-1}XK_2]\nonumber\\
&=&(K_1^{-1})'XK_2+K_1^{-1}X'K_2+K_1^{-1}XK_2'\nonumber\\
&=&-K_1^{-1}K_1'K_1^{-1}XK_2+K_1^{-1}X'K_2+K_1^{-1}XK_2'\nonumber\\
&=&K_1^{-1}X [-X^{-1}K_1'K_1^{-1}XK_2+X^{-1}X'K_2+K_2']\nonumber\\
&=&K_1^{-1}X [X^{-1}G^{(1)}XK_2+X^{-1}X'K_2+K_2']\label{XG}\\
&=&0, \nonumber
\end{eqnarray}
where Eq. (\ref{XG}) is valid due to Eq. (\ref{K12X}).

A direct consequence of Eq. (\ref{d1}) is, for any $s$, we have
\begin{eqnarray}
\tilde{K}_j^{(2)}(s)X(s)=\tilde{K}_j^{(1)}(s)X(0). \label{K12}
\end{eqnarray}
Now we have
\begin{eqnarray}
K_j^{(2)}(s)u_m(s)=\sum_n X_{nm}(0) K_j^{(1)}(s)v_n(s).\label{Duv}
\end{eqnarray}
In fact, one can verify that
\begin{eqnarray*}
&&K_j^{(2)}(s)u_m(s)\\
&=&\sum_i(\tilde{K}_j^{(2)})_{im}(s)u_i(s)\\
&=&\sum_i(\tilde{K}_j^{(2)})_{im}(s)\sum_l X_{li}(s)v_l(s)\\
&=&\sum_l [(\tilde{K}_j^{(2)})_{im}(s) X_{li}(s)]v_l(s)\\
&=&\sum_l [\tilde{K}_j^{(1)}(s)X(0)]_{lm}v_l(s)\\
&=&\sum_l \sum_n(\tilde{K}_j^{(1)})_{ln}X_{nm}(0)v_l(s)\\
&=&\sum_n X_{nm}(0) K_j^{(1)}(s)v_n(s).
\end{eqnarray*}

Given some eigenstate $\phi(0)$ (unnormalised for convenience), $\phi(0)=\sum_m \phi_m^{(1)}(0)v_m(0)$, where $m=j_1,j_2,...,j_h$. Now according to Eq. (\ref{adiabatic0}), one can see that $\phi^{(1)}(s)=\sum_m \phi_m^{(1)}(0)K_j^{(1)}(s)v_m(s)$. Similarly, $\phi(0)=\sum_m \phi_m^{(2)}(0)u_m(0)$,
$\phi^{(2)}(s)=\sum_m \phi_m^{(2)}(0)K_j^{(2)}(s)u_m(s)$.
Note that since $u_i(s)=\sum_j X_{ji}(s)v_j(s)$ $\sum_m \phi_m^{(2)}(0) X_{nm}(0)=\phi_n^{(1)}(0)$. According to Eq. (\ref{Duv}),
\begin{eqnarray}
&&\phi^{(2)}(s)\nonumber\\
&=&\sum_m \phi_m^{(2)}(0)K_j^{(2)}(s)u_m(s)\nonumber\\
&=&\sum_m \phi_m^{(2)}(0)\sum_n X_{nm}(0) K_j^{(1)}(s)v_n(s)\nonumber\\
&=&\sum_n \phi_n^{(1)}(0)K_j^{(1)}(s)v_n(s)\nonumber\\
&=&\phi^{(1)}(s),
\end{eqnarray}
which shows the evolution is unique, regardless of the specific choice of the basis.

In the non-degenerate case, the above result can be shown in an explicit way.
Let $v_j(s)$ be the eigenvectors of $H(s)$ and $\xi_j(s)$ form a biorthogonal system. Assume that there is another choice of eigenvectors $\psi_j(s)$ and $\phi_i(s)$ form a biorthogonal system.  Now since $H(s)$ has nondegenerate eigenvalues,
 it is obvious that $\psi_j(s)=\mu_j(s)v_j(s)$ and $\phi_j(s)=\frac{1}{\overline{\mu}_j(s)}\xi_j(s)$, where
$\mu_j(s)$ is some complex number and $\overline{\mu}_j(s)$ is its conjugate.

Now we consider the relation of $v_j(s)e^{i\int_0^s i\braket{\xi_j(\sigma)|\dot{v}_j(\sigma)}d\sigma}$ and
$\psi_j(s)e^{i\int_0^s i\braket{\phi_j(\sigma)|\dot{\psi}_j(\sigma)}d\sigma}$. In fact,
\begin{eqnarray*}
&&\psi_j(s)e^{i\int_0^s i\braket{\phi_j(\sigma)|\dot{\psi}_j(\sigma)}d\sigma}\\
&=&\mu_j(s)v_j(s)e^{-\int_0^s \braket{\phi_j(\sigma)|\dot{\psi}_j(\sigma)}d\sigma}\\
&=&\mu_j(s)v_j(s)e^{-\int_0^s \frac{1}{\mu_j(s)}\braket{\xi_j(\sigma)|[\mu_j(\sigma)v_j(\sigma)]'}d\sigma}\\
&=&\mu_j(s)v_j(s)e^{-\int_0^s \frac{\dot{\mu}_j(s)}{\mu_j(s)}d\sigma -\braket{\xi_j(\sigma)|\dot{v}_j(\sigma)}d\sigma}\\
&=&\mu_j(0)v_j(s)e^{-\int_0^s \braket{\xi_j(\sigma)|\dot{v}_j(\sigma)}d\sigma}.
\end{eqnarray*}

Now suppose that one has an initial state (unnormalized vector) $q_j(0)=c_j(0)v_j(0)$.
The above derivation and direct calculations
show that the adiabatically evolved state $q_j(s)$ can always be written as
$q_j(s)=c_j(0)v_j(s)e^{-\int_0^s \braket{\xi_j(\sigma)|\dot{v}_j(\sigma)}d\sigma}$
 regardless of using $v_j$ or $\psi_j$ as the reference in calculation.
It should also be mentioned that the Berry phase in our discussion is complex, which is similar to \cite{garrison1988complex}. In fact, a complex Berry phase and the biorthogonal states are also used in \cite{leclerc2012role}. The non-Abelian geometric phase in also discussed in the non-adiabatic case in \cite{lou2025nonadiabatic}.

\subsection{The uniform boundedness of $U_T(s)$ and $U_T^{-1}(s)$}
We still have to prove $U_T(s)$ and $U_T^{-1}(s)$ are uniformly bounded with respect to $T$.

For $U_T^{-1}(s)$, one can take the adjoint of the Eq. (\ref{zero}), then
\begin{eqnarray}
\nonumber&&\frac{d}{ds}[\tilde{v}^\dag_{j_l}(s) e^{iT\int_0^s\lambda_j(\sigma)d\sigma}[U_T^{-1}]^\dag(s)]\\
&=& \frac{d}{ds}[\tilde{v}^\dag_{j_l}(s)] e^{iT\int_0^s\lambda_j(\sigma)d\sigma}[U_T^{-1}]^\dag(s).\label{zero2}
\end{eqnarray}
Construct a matrix $\tilde{V}(s)$, whose column vectors are $\tilde{v}_{j_l}(s)$ and denote $E_T(s)$ the diagonal matrix
whose diagonal entries are $e^{iT\int_0^s\lambda_j(\sigma)d\sigma}$. It follows from
Eq. (\ref{zero2}) that
\begin{eqnarray}
\nonumber&&\frac{d}{ds}[E_T(s)\tilde{V}^\dag(s) [U_T^{-1}]^\dag(s)]\\
&=& E_T(s)\frac{d}{ds}[\tilde{V}^\dag(s)] [U_T^{-1}]^\dag(s).\label{zero3}
\end{eqnarray}
Eq. (\ref{zero3}) can be rewritten as
\begin{eqnarray}
\nonumber&&\frac{d}{ds}[E_T(s)\tilde{V}^\dag(s) [U_T^{-1}]^\dag(s)]\\
\nonumber&=& E_T(s)\frac{d}{ds}[\tilde{V}^\dag(s)][\tilde{V}^\dag(s)]^{-1} E^\dag_T(s)[E_T(s)\tilde{V}^\dag(s)[U_T^{-1}]^\dag(s)]\label{zero4}
\end{eqnarray}

By taking integral,
\begin{eqnarray}
\nonumber&&E_T(s)\tilde{V}^\dag(s)[U_T^{-1}]^\dag(s)-\tilde{V}^\dag(0)\\
\nonumber&=& \int_0^s E_T(\sigma)\frac{d}{d\sigma} [\tilde{V}^\dag(\sigma)][\tilde{V}^\dag(\sigma)]^{-1} E^\dag_T(\sigma)\\
\nonumber&& \cdot[E_T(\sigma)\tilde{V}^\dag(\sigma)[U_T^{-1}]^\dag(\sigma)]d\sigma.\\
\end{eqnarray}

By using the triangle inequality of norm and Gr\"{o}nwall inequality (see Appendix B), one can see that
\begin{eqnarray}
\nonumber &&\|E_T(s)\tilde{V}^\dag(s)[U_T^{-1}]^\dag(s)\|\leqslant \|\tilde{V}^\dag(0)\|\\
\nonumber &+&\int_0^s \|\tilde{V}^\dag(0)\|\|E_T(\sigma)\frac{d}{d\sigma} [\tilde{V}^\dag(\sigma)][\tilde{V}^\dag(\sigma)]^{-1} E^\dag_T(\sigma)\|\\
\nonumber &&\cdot e^{\int_\sigma^s \|E_T(r)\frac{d}{dr} [\tilde{V}^\dag(r)][\tilde{V}^\dag(r)]^{-1}E^\dag_T(r)\|dr} d\sigma.\\
\label{norm1}
\end{eqnarray}
Note that both $\tilde{V}(s)$ and $\frac{d}{ds} [\tilde{V}^\dag(s)][\tilde{V}^\dag(s)]^{-1}$ are continuous and independent of $T$, thus they are uniformly bounded. $E_T(s)$ is also uniformly bounded.
It follows from Eq. (\ref{norm1}) that $\|E_T(s)\tilde{V}^\dag(s)[U_T^{-1}]^\dag(s)\|$ is uniformly bounded.  Thus $U^{-1}_T(s)$ is uniformly bounded with respect to $T$.

Along similar lines, one can prove that $U_T(s)$ is also uniformly bounded.
In fact,
\begin{eqnarray}
\nonumber&&\frac{d}{ds}[\xi^\dag_{j_l}(s)e^{iT\int_0^s\lambda_j(\sigma)d\sigma}U_T(s)]\\
\nonumber&=&\frac{d}{ds}[\xi^\dag_{j_l}(s)]e^{iT\int_0^s\lambda_j(\sigma)d\sigma}U_T(s)\\
\nonumber&+&\xi^\dag_{j_l}(s)\frac{d}{ds}[e^{iT\int_0^s\lambda_j(\sigma)d\sigma}U_T(s)]
\end{eqnarray}
Note that by Eqs. (\ref{resol1}) and (\ref{resol2}),
\begin{eqnarray}
\nonumber&&\xi^\dag_{j_l}(s)\frac{d}{ds}[e^{iT\int_0^s\lambda_j(\sigma)d\sigma}U_T(s)]\\
\nonumber&=&\xi^\dag_{j_l}(s) iT e^{iT\int_0^s\lambda_j(\sigma)d\sigma}[\lambda_j(s)I-H(s)]U_T(s)\\
\nonumber&=&0.
\end{eqnarray}
Hence it follows that
\begin{eqnarray}
\nonumber&&\frac{d}{ds}[\xi^\dag_{j_l}(s)e^{iT\int_0^s\lambda_j(\sigma)d\sigma}U_T(s)]\\
\nonumber&=&\frac{d}{ds}[\xi^\dag_{j_l}(s)]e^{iT\int_0^s\lambda_j(\sigma)d\sigma}U_T(s).\label{zero5}\\
\end{eqnarray}

Construct a matrix $\Xi(s)$, whose column vectors are $\xi_{j_l}(s)$. It follows from Eq. (\ref{zero5}) that

\begin{eqnarray}
\nonumber&&\frac{d}{ds}[E_T(s)\Xi^\dag(s)U_T(s)]\\
\nonumber&=&E_T(s)\frac{d}{ds}[\Xi^\dag(s)]U_T(s)\\
\nonumber&=&E_T(s)\frac{d}{ds}[\Xi^\dag(s)][\Xi^\dag(s)]^{-1}E^\dag_T(s)[E_T(s)\Xi^\dag(s)U_T(s)]\label{zero6}\\
\end{eqnarray}
Applying the Gr\"{o}nwall inequality,
\begin{eqnarray}
\nonumber &&\|E_T(s)\Xi^\dag(s)U_T(s)\|\leqslant \|\Xi^\dag(0)\|\\
\nonumber &+&\int_0^s \|\Xi^\dag(0)\|\|E_T(\sigma)\frac{d}{d\sigma} [\Xi^\dag(\sigma)][\Xi^\dag(\sigma)]^{-1}E^\dag_T(\sigma)\|\\
\nonumber &&\cdot e^{\int_\sigma^s \|E_T(r)\frac{d}{dr} [\Xi^\dag(r)][\Xi^\dag(r)]^{-1}E^\dag_T(r)\|dr} d\sigma.\\
\label{norm2}
\end{eqnarray}
Since $\Xi(s)$ is continuous and independent of $T$, $E_T(s)$ is unitary,  a similar discussion shows that $U_T(s)$ is uniformly bounded with respect to $T$.

\section{Discussions and Conclusion}
The above discussion shows that the adiabatic theorem is valid for the diagonalizable Hamiltonians with real eigenvalues.
That is, the eigenstate will remain in the subspace when evolving slowly enough. When the energy is non-degenerate, it reduces
to the Berry phase.

Apparently, the result is somewhat natural. One may wonder why such a result deserves a specified proof. In fact,
any Hamiltonian $H(s)$ with non-degenerate real eigenvalues is similar to a Hermitian Hamiltonian $H_2(s)$. If
$H(s)=V^{-1}H_2(s)V$, then the evolution operator of $U_T(s)=V^{-1}U^{(2)}_T(s)$, where $U^{(2)}_T(s)$ is the evolution operator
of $TH_2(s)$. Moreover, if $x(s)$ is an eigenvector of $H_2(s)$, then $V^{-1}x(s)$ is an eigenvector of $H(s)$.
Now since $H_2(s)$ is Hermitian, the adiabatic theorem is valid for $H_2(s)$. Moreover, since $V$ is time independent and $U_T(s)=V^{-1}U^{(2)}_T(s)$, one can directly verify that the adiabatic theorem is also valid for $H(s)$. However, when  $H(s)=V^{-1}(s)H_2(s)V(s)$, i.e. the matrix $V(s)$ is dependent of $s$, one has $U_T(s)\neq V^{-1}(s)U^{(2)}_T(s)$. In this case, one cannot directly obtain the adiabatic theorem for $H(s)$ from $H_2(s)$. Hence a proof is necessary.

Unlike the Hermitian case, the eigenvectors of the Hamiltonian $H(s)$ are usually unnormalized in the non-Hermitian case for
convenience. That is, the eigenvectors are not uniquely defined since their norms are undetermined. Consequently, the Berry phases are not uniquely determined either. However, this will bring no essential change to the result. As we have seen in Section III B, this is
even true in the non-Abelian case.

 In Eqs. (\ref{S})-(\ref{s2}), our proof utilizes the spectral decomposition and functional calculus with respect to biorthogonal systems. Such an approach is inspired by Kato's discussions on Hermitian systems, which can avoid the Riemann-Lebesgue Lemma in the
 discussion \cite{kato1950adiabatic}. However, since its utilization of orthogonal projections and unitary matrices, Kato's discussions do not
directly apply to the non-Hermitian case. Moreover, without the Berry phase or its non-abelian counterpart,
 a treatment of the differential equation of the orthogonal projections and the intertwining operator is necessary in the proof.
  This long procedure is not needed in this paper. Due to the use of Berry phase or the non-abelian geometric phase, the proof is largely simplified.

There is one place where this paper is more complex than the Hermitian case. One has to show the uniform boundedness of $U_T(s)$ and $U_T^{-1}(s)$, which are obvious in Hermitian quantum mechanics. To show the uniform boundedness, we use the Gr\"{o}nwall inequality. Given that our discussion are in a more general case, this complication is not unexpected.

In conclusion, we obtained a more general adiabatic theorem for non-Hermitian systems with real eigenvalues. The discussions also partly justifies the definition of a complex Berry phase and its non-abelian generalization.

\section{Appendix}

\subsection{The Gr\"{o}nwall inequality}

The integral form of the Gr\"{o}nwall inequality is as follows \cite{gronwall1919note,bellman1943stability}.
Let $I=[a,b)$ be an interval. If $\beta$ is non-negative and $u$ satisfies the following
inequality,
\[
u(t)\leqslant \alpha(t)+\int_a^t \beta(s)u(s)ds, \forall t\in I,
\]
then
\[
u(t)\leqslant \alpha(t)+\int_a^t \alpha(s)\beta(s)e^{\int_s^t \beta(r)dr}ds.
\]


\section*{Acknowledgement}
This work is partially supported by the National Natural Science Foundation of China (12371135).
The first author thanks Prof. Chunlan Jiang, Dianmin Tong and Zeqian Chen for useful suggestions.

\begin{thebibliography}{32}%
\makeatletter
\providecommand \@ifxundefined [1]{%
 \@ifx{#1\undefined}
}%
\providecommand \@ifnum [1]{%
 \ifnum #1\expandafter \@firstoftwo
 \else \expandafter \@secondoftwo
 \fi
}%
\providecommand \@ifx [1]{%
 \ifx #1\expandafter \@firstoftwo
 \else \expandafter \@secondoftwo
 \fi
}%
\providecommand \natexlab [1]{#1}%
\providecommand \enquote  [1]{``#1''}%
\providecommand \bibnamefont  [1]{#1}%
\providecommand \bibfnamefont [1]{#1}%
\providecommand \citenamefont [1]{#1}%
\providecommand \href@noop [0]{\@secondoftwo}%
\providecommand \href [0]{\begingroup \@sanitize@url \@href}%
\providecommand \@href[1]{\@@startlink{#1}\@@href}%
\providecommand \@@href[1]{\endgroup#1\@@endlink}%
\providecommand \@sanitize@url [0]{\catcode `\\12\catcode `\$12\catcode
  `\&12\catcode `\#12\catcode `\^12\catcode `\_12\catcode `\%12\relax}%
\providecommand \@@startlink[1]{}%
\providecommand \@@endlink[0]{}%
\providecommand \url  [0]{\begingroup\@sanitize@url \@url }%
\providecommand \@url [1]{\endgroup\@href {#1}{\urlprefix }}%
\providecommand \urlprefix  [0]{URL }%
\providecommand \Eprint [0]{\href }%
\providecommand \doibase [0]{http://dx.doi.org/}%
\providecommand \selectlanguage [0]{\@gobble}%
\providecommand \bibinfo  [0]{\@secondoftwo}%
\providecommand \bibfield  [0]{\@secondoftwo}%
\providecommand \translation [1]{[#1]}%
\providecommand \BibitemOpen [0]{}%
\providecommand \bibitemStop [0]{}%
\providecommand \bibitemNoStop [0]{.\EOS\space}%
\providecommand \EOS [0]{\spacefactor3000\relax}%
\providecommand \BibitemShut  [1]{\csname bibitem#1\endcsname}%
\let\auto@bib@innerbib\@empty
\bibitem [{\citenamefont {Born}\ and\ \citenamefont
  {Fock}(1928)}]{born1928beweis}%
  \BibitemOpen
  \bibfield  {author} {\bibinfo {author} {\bibfnamefont {M.}~\bibnamefont
  {Born}}\ and\ \bibinfo {author} {\bibfnamefont {V.}~\bibnamefont {Fock}},\
  }\href@noop {} {\bibfield  {journal} {\bibinfo  {journal} {Zeitschrift
  f{\"u}r Physik}\ }\textbf {\bibinfo {volume} {51}},\ \bibinfo {pages} {165}
  (\bibinfo {year} {1928})}\BibitemShut {NoStop}%
\bibitem [{\citenamefont {Kato}(1950)}]{kato1950adiabatic}%
  \BibitemOpen
  \bibfield  {author} {\bibinfo {author} {\bibfnamefont {T.}~\bibnamefont
  {Kato}},\ }\href@noop {} {\bibfield  {journal} {\bibinfo  {journal} {Journal
  of the Physical Society of Japan}\ }\textbf {\bibinfo {volume} {5}},\
  \bibinfo {pages} {435} (\bibinfo {year} {1950})}\BibitemShut {NoStop}%
\bibitem [{\citenamefont {Avron}\ and\ \citenamefont
  {Elgart}(1999)}]{avron1999adiabatic}%
  \BibitemOpen
  \bibfield  {author} {\bibinfo {author} {\bibfnamefont {J.~E.}\ \bibnamefont
  {Avron}}\ and\ \bibinfo {author} {\bibfnamefont {A.}~\bibnamefont {Elgart}},\
  }\href@noop {} {\bibfield  {journal} {\bibinfo  {journal} {Communications in
  mathematical physics}\ }\textbf {\bibinfo {volume} {203}},\ \bibinfo {pages}
  {445} (\bibinfo {year} {1999})}\BibitemShut {NoStop}%
\bibitem [{\citenamefont {Teufel}(2001)}]{teufel2001note}%
  \BibitemOpen
  \bibfield  {author} {\bibinfo {author} {\bibfnamefont {S.}~\bibnamefont
  {Teufel}},\ }\href@noop {} {\bibfield  {journal} {\bibinfo  {journal}
  {Letters in Mathematical Physics}\ }\textbf {\bibinfo {volume} {58}},\
  \bibinfo {pages} {261} (\bibinfo {year} {2001})}\BibitemShut {NoStop}%
\bibitem [{\citenamefont {Tong}(2010)}]{tong2010quantitative}%
  \BibitemOpen
  \bibfield  {author} {\bibinfo {author} {\bibfnamefont {D.}~\bibnamefont
  {Tong}},\ }\href@noop {} {\bibfield  {journal} {\bibinfo  {journal} {Physical
  review letters}\ }\textbf {\bibinfo {volume} {104}},\ \bibinfo {pages}
  {120401} (\bibinfo {year} {2010})}\BibitemShut {NoStop}%
\bibitem [{\citenamefont {Berry}(1984)}]{berry1984quantal}%
  \BibitemOpen
  \bibfield  {author} {\bibinfo {author} {\bibfnamefont {M.~V.}\ \bibnamefont
  {Berry}},\ }\href@noop {} {\bibfield  {journal} {\bibinfo  {journal}
  {Proceedings of the Royal Society of London. A. Mathematical and Physical
  Sciences}\ }\textbf {\bibinfo {volume} {392}},\ \bibinfo {pages} {45}
  (\bibinfo {year} {1984})}\BibitemShut {NoStop}%
\bibitem [{\citenamefont {Farhi}\ \emph {et~al.}(2001)\citenamefont {Farhi},
  \citenamefont {Goldstone}, \citenamefont {Gutmann}, \citenamefont {Lapan},
  \citenamefont {Lundgren},\ and\ \citenamefont {Preda}}]{farhi2001quantum}%
  \BibitemOpen
  \bibfield  {author} {\bibinfo {author} {\bibfnamefont {E.}~\bibnamefont
  {Farhi}}, \bibinfo {author} {\bibfnamefont {J.}~\bibnamefont {Goldstone}},
  \bibinfo {author} {\bibfnamefont {S.}~\bibnamefont {Gutmann}}, \bibinfo
  {author} {\bibfnamefont {J.}~\bibnamefont {Lapan}}, \bibinfo {author}
  {\bibfnamefont {A.}~\bibnamefont {Lundgren}}, \ and\ \bibinfo {author}
  {\bibfnamefont {D.}~\bibnamefont {Preda}},\ }\href@noop {} {\bibfield
  {journal} {\bibinfo  {journal} {Science}\ }\textbf {\bibinfo {volume}
  {292}},\ \bibinfo {pages} {472} (\bibinfo {year} {2001})}\BibitemShut
  {NoStop}%
\bibitem [{\citenamefont {Albash}\ and\ \citenamefont
  {Lidar}(2018)}]{albash2018adiabatic}%
  \BibitemOpen
  \bibfield  {author} {\bibinfo {author} {\bibfnamefont {T.}~\bibnamefont
  {Albash}}\ and\ \bibinfo {author} {\bibfnamefont {D.~A.}\ \bibnamefont
  {Lidar}},\ }\href@noop {} {\bibfield  {journal} {\bibinfo  {journal} {Reviews
  of Modern Physics}\ }\textbf {\bibinfo {volume} {90}},\ \bibinfo {pages}
  {015002} (\bibinfo {year} {2018})}\BibitemShut {NoStop}%
\bibitem [{\citenamefont {Wilczek}\ and\ \citenamefont
  {Zee}(1984)}]{wilczek1984appearance}%
  \BibitemOpen
  \bibfield  {author} {\bibinfo {author} {\bibfnamefont {F.}~\bibnamefont
  {Wilczek}}\ and\ \bibinfo {author} {\bibfnamefont {A.}~\bibnamefont {Zee}},\
  }\href@noop {} {\bibfield  {journal} {\bibinfo  {journal} {Physical Review
  Letters}\ }\textbf {\bibinfo {volume} {52}},\ \bibinfo {pages} {2111}
  (\bibinfo {year} {1984})}\BibitemShut {NoStop}%
\bibitem [{\citenamefont {Bender}\ and\ \citenamefont
  {Boettcher}(1998)}]{bender1998real}%
  \BibitemOpen
  \bibfield  {author} {\bibinfo {author} {\bibfnamefont {C.~M.}\ \bibnamefont
  {Bender}}\ and\ \bibinfo {author} {\bibfnamefont {S.}~\bibnamefont
  {Boettcher}},\ }\href@noop {} {\bibfield  {journal} {\bibinfo  {journal}
  {Phys. Rev. Lett.}\ }\textbf {\bibinfo {volume} {80}},\ \bibinfo {pages}
  {5243} (\bibinfo {year} {1998})}\BibitemShut {NoStop}%
\bibitem [{\citenamefont {Bender}(2007)}]{bender2007making}%
  \BibitemOpen
  \bibfield  {author} {\bibinfo {author} {\bibfnamefont {C.~M.}\ \bibnamefont
  {Bender}},\ }\href@noop {} {\bibfield  {journal} {\bibinfo  {journal} {Rep.
  Prog. Phys.}\ }\textbf {\bibinfo {volume} {70}},\ \bibinfo {pages} {947}
  (\bibinfo {year} {2007})}\BibitemShut {NoStop}%
\bibitem [{\citenamefont {Mostafazadeh}(2010)}]{mostafazadeh2010pseudo}%
  \BibitemOpen
  \bibfield  {author} {\bibinfo {author} {\bibfnamefont {A.}~\bibnamefont
  {Mostafazadeh}},\ }\href@noop {} {\bibfield  {journal} {\bibinfo  {journal}
  {Int. J. Geom. Methods Mod. Phys.}\ }\textbf {\bibinfo {volume} {7}},\
  \bibinfo {pages} {1191} (\bibinfo {year} {2010})}\BibitemShut {NoStop}%
\bibitem [{\citenamefont {Ashida}\ \emph {et~al.}(2020)\citenamefont {Ashida},
  \citenamefont {Gong},\ and\ \citenamefont {Ueda}}]{ashida2020non}%
  \BibitemOpen
  \bibfield  {author} {\bibinfo {author} {\bibfnamefont {Y.}~\bibnamefont
  {Ashida}}, \bibinfo {author} {\bibfnamefont {Z.}~\bibnamefont {Gong}}, \ and\
  \bibinfo {author} {\bibfnamefont {M.}~\bibnamefont {Ueda}},\ }\href@noop {}
  {\bibfield  {journal} {\bibinfo  {journal} {Advances in Physics}\ }\textbf
  {\bibinfo {volume} {69}},\ \bibinfo {pages} {249} (\bibinfo {year}
  {2020})}\BibitemShut {NoStop}%
\bibitem [{\citenamefont {Yao}\ and\ \citenamefont
  {Wang}(2018)}]{PhysRevLett.121.086803}%
  \BibitemOpen
  \bibfield  {author} {\bibinfo {author} {\bibfnamefont {S.}~\bibnamefont
  {Yao}}\ and\ \bibinfo {author} {\bibfnamefont {Z.}~\bibnamefont {Wang}},\
  }{\bibfield  {journal}
  {\bibinfo  {journal} {Phys. Rev. Lett.}\ }\textbf {\bibinfo {volume} {121}},\
  \bibinfo {pages} {086803} (\bibinfo {year} {2018})}\BibitemShut {NoStop}%
\bibitem [{\citenamefont {Yao}\ \emph {et~al.}(2018)\citenamefont {Yao},
  \citenamefont {Song},\ and\ \citenamefont {Wang}}]{PhysRevLett.121.136802}%
  \BibitemOpen
  \bibfield  {author} {\bibinfo {author} {\bibfnamefont {S.}~\bibnamefont
  {Yao}}, \bibinfo {author} {\bibfnamefont {F.}~\bibnamefont {Song}}, \ and\
  \bibinfo {author} {\bibfnamefont {Z.}~\bibnamefont {Wang}},\ }{\bibfield  {journal} {\bibinfo  {journal}
  {Phys. Rev. Lett.}\ }\textbf {\bibinfo {volume} {121}},\ \bibinfo {pages}
  {136802} (\bibinfo {year} {2018})}\BibitemShut {NoStop}%
\bibitem [{\citenamefont {Alvarez}\ \emph {et~al.}(2018)\citenamefont
  {Alvarez}, \citenamefont {Vargas},\ and\ \citenamefont
  {Torres}}]{alvarez2018non}%
  \BibitemOpen
  \bibfield  {author} {\bibinfo {author} {\bibfnamefont {V.~M.}\ \bibnamefont
  {Alvarez}}, \bibinfo {author} {\bibfnamefont {J.~B.}\ \bibnamefont {Vargas}},
  \ and\ \bibinfo {author} {\bibfnamefont {L.~F.}\ \bibnamefont {Torres}},\
  }\href@noop {} {\bibfield  {journal} {\bibinfo  {journal} {Physical Review
  B}\ }\textbf {\bibinfo {volume} {97}},\ \bibinfo {pages} {121401} (\bibinfo
  {year} {2018})}\BibitemShut {NoStop}%
\bibitem [{\citenamefont {{\"O}zdemir}\ \emph {et~al.}(2019)\citenamefont
  {{\"O}zdemir}, \citenamefont {Rotter}, \citenamefont {Nori},\ and\
  \citenamefont {Yang}}]{ozdemir2019parity}%
  \BibitemOpen
  \bibfield  {author} {\bibinfo {author} {\bibfnamefont {{\c{S}}.~K.}\
  \bibnamefont {{\"O}zdemir}}, \bibinfo {author} {\bibfnamefont
  {S.}~\bibnamefont {Rotter}}, \bibinfo {author} {\bibfnamefont
  {F.}~\bibnamefont {Nori}}, \ and\ \bibinfo {author} {\bibfnamefont
  {L.}~\bibnamefont {Yang}},\ }\href@noop {} {\bibfield  {journal} {\bibinfo
  {journal} {Nature materials}\ }\textbf {\bibinfo {volume} {18}},\ \bibinfo
  {pages} {783} (\bibinfo {year} {2019})}\BibitemShut {NoStop}%
\bibitem [{\citenamefont {Song}\ \emph
  {et~al.}(2019{\natexlab{a}})\citenamefont {Song}, \citenamefont {Yao},\ and\
  \citenamefont {Wang}}]{PhysRevLett.123.170401}%
  \BibitemOpen
  \bibfield  {author} {\bibinfo {author} {\bibfnamefont {F.}~\bibnamefont
  {Song}}, \bibinfo {author} {\bibfnamefont {S.}~\bibnamefont {Yao}}, \ and\
  \bibinfo {author} {\bibfnamefont {Z.}~\bibnamefont {Wang}},\ } {\bibfield  {journal} {\bibinfo  {journal}
  {Phys. Rev. Lett.}\ }\textbf {\bibinfo {volume} {123}},\ \bibinfo {pages}
  {170401} (\bibinfo {year} {2019}{\natexlab{a}})}\BibitemShut {NoStop}%
\bibitem [{\citenamefont {Song}\ \emph
  {et~al.}(2019{\natexlab{b}})\citenamefont {Song}, \citenamefont {Yao},\ and\
  \citenamefont {Wang}}]{PhysRevLett.123.246801}%
  \BibitemOpen
  \bibfield  {author} {\bibinfo {author} {\bibfnamefont {F.}~\bibnamefont
  {Song}}, \bibinfo {author} {\bibfnamefont {S.}~\bibnamefont {Yao}}, \ and\
  \bibinfo {author} {\bibfnamefont {Z.}~\bibnamefont {Wang}},\ } {\bibfield  {journal} {\bibinfo  {journal}
  {Phys. Rev. Lett.}\ }\textbf {\bibinfo {volume} {123}},\ \bibinfo {pages}
  {246801} (\bibinfo {year} {2019}{\natexlab{b}})}\BibitemShut {NoStop}%
\bibitem [{\citenamefont {Longhi}(2019)}]{PhysRevResearch.1.023013}%
  \BibitemOpen
  \bibfield  {author} {\bibinfo {author} {\bibfnamefont {S.}~\bibnamefont
  {Longhi}},\ }{\bibfield
  {journal} {\bibinfo  {journal} {Phys. Rev. Research}\ }\textbf {\bibinfo
  {volume} {1}},\ \bibinfo {pages} {023013} (\bibinfo {year}
  {2019})}\BibitemShut {NoStop}%
\bibitem [{\citenamefont {Garrison}\ and\ \citenamefont
  {Wright}(1988)}]{garrison1988complex}%
  \BibitemOpen
  \bibfield  {author} {\bibinfo {author} {\bibfnamefont {J.}~\bibnamefont
  {Garrison}}\ and\ \bibinfo {author} {\bibfnamefont {E.~M.}\ \bibnamefont
  {Wright}},\ }\href@noop {} {\bibfield  {journal} {\bibinfo  {journal}
  {Physics Letters A}\ }\textbf {\bibinfo {volume} {128}},\ \bibinfo {pages}
  {177} (\bibinfo {year} {1988})}\BibitemShut {NoStop}%
\bibitem [{\citenamefont {Nenciu}\ and\ \citenamefont
  {Rasche}(1992)}]{nenciu1992adiabatic}%
  \BibitemOpen
  \bibfield  {author} {\bibinfo {author} {\bibfnamefont {G.}~\bibnamefont
  {Nenciu}}\ and\ \bibinfo {author} {\bibfnamefont {G.}~\bibnamefont
  {Rasche}},\ }\href@noop {} {\bibfield  {journal} {\bibinfo  {journal}
  {Journal of Physics A: Mathematical and General}\ }\textbf {\bibinfo {volume}
  {25}},\ \bibinfo {pages} {5741} (\bibinfo {year} {1992})}\BibitemShut
  {NoStop}%
\bibitem [{\citenamefont {Mehri-Dehnavi}\ and\ \citenamefont
  {Mostafazadeh}(2008)}]{mehri2008geometric}%
  \BibitemOpen
  \bibfield  {author} {\bibinfo {author} {\bibfnamefont {H.}~\bibnamefont
  {Mehri-Dehnavi}}\ and\ \bibinfo {author} {\bibfnamefont {A.}~\bibnamefont
  {Mostafazadeh}},\ }\href@noop {} {\bibfield  {journal} {\bibinfo  {journal}
  {Journal of Mathematical Physics}\ }\textbf {\bibinfo {volume} {49}}
  (\bibinfo {year} {2008})}\BibitemShut {NoStop}%
\bibitem [{\citenamefont {Berry}\ and\ \citenamefont
  {Uzdin}(2011)}]{berry2011slow}%
  \BibitemOpen
  \bibfield  {author} {\bibinfo {author} {\bibfnamefont {M.}~\bibnamefont
  {Berry}}\ and\ \bibinfo {author} {\bibfnamefont {R.}~\bibnamefont {Uzdin}},\
  }\href@noop {} {\bibfield  {journal} {\bibinfo  {journal} {Journal of Physics
  A: Mathematical and Theoretical}\ }\textbf {\bibinfo {volume} {44}},\
  \bibinfo {pages} {435303} (\bibinfo {year} {2011})}\BibitemShut {NoStop}%
\bibitem [{\citenamefont {Jing}\ \emph {et~al.}(2016)\citenamefont {Jing},
  \citenamefont {Sarandy}, \citenamefont {Lidar}, \citenamefont {Luo},\ and\
  \citenamefont {Wu}}]{jing2016eigenstate}%
  \BibitemOpen
  \bibfield  {author} {\bibinfo {author} {\bibfnamefont {J.}~\bibnamefont
  {Jing}}, \bibinfo {author} {\bibfnamefont {M.~S.}\ \bibnamefont {Sarandy}},
  \bibinfo {author} {\bibfnamefont {D.~A.}\ \bibnamefont {Lidar}}, \bibinfo
  {author} {\bibfnamefont {D.-W.}\ \bibnamefont {Luo}}, \ and\ \bibinfo
  {author} {\bibfnamefont {L.-A.}\ \bibnamefont {Wu}},\ }\href@noop {}
  {\bibfield  {journal} {\bibinfo  {journal} {Physical Review A}\ }\textbf
  {\bibinfo {volume} {94}},\ \bibinfo {pages} {042131} (\bibinfo {year}
  {2016})}\BibitemShut {NoStop}%
\bibitem [{\citenamefont {Gong}\ \emph {et~al.}(2019)\citenamefont {Gong},
  \citenamefont {Wang} \emph {et~al.}}]{gong2019piecewise}%
  \BibitemOpen
  \bibfield  {author} {\bibinfo {author} {\bibfnamefont {J.}~\bibnamefont
  {Gong}}, \bibinfo {author} {\bibfnamefont {Q.-h.}\ \bibnamefont {Wang}},
  \emph {et~al.},\ }\href@noop {} {\bibfield  {journal} {\bibinfo  {journal}
  {Physical Review A}\ }\textbf {\bibinfo {volume} {99}},\ \bibinfo {pages}
  {012107} (\bibinfo {year} {2019})}\BibitemShut {NoStop}%
\bibitem [{\citenamefont {Jin}\ and\ \citenamefont
  {Jing}(2025)}]{jin2025universal}%
  \BibitemOpen
  \bibfield  {author} {\bibinfo {author} {\bibfnamefont {Z.-y.}\ \bibnamefont
  {Jin}}\ and\ \bibinfo {author} {\bibfnamefont {J.}~\bibnamefont {Jing}},\
  }\href@noop {} {\bibfield  {journal} {\bibinfo  {journal} {Physical Review
  A}\ }\textbf {\bibinfo {volume} {112}},\ \bibinfo {pages} {032605} (\bibinfo
  {year} {2025})}\BibitemShut {NoStop}%
\bibitem [{\citenamefont {Kumar}\ \emph {et~al.}(2025)\citenamefont {Kumar},
  \citenamefont {Gefen},\ and\ \citenamefont {Snizhko}}]{kumar2025general}%
  \BibitemOpen
  \bibfield  {author} {\bibinfo {author} {\bibfnamefont {P.}~\bibnamefont
  {Kumar}}, \bibinfo {author} {\bibfnamefont {Y.}~\bibnamefont {Gefen}}, \ and\
  \bibinfo {author} {\bibfnamefont {K.}~\bibnamefont {Snizhko}},\ }\href@noop
  {} {\bibfield  {journal} {\bibinfo  {journal} {arXiv preprint
  arXiv:2502.04214}\ } (\bibinfo {year} {2025})}\BibitemShut {NoStop}%
\bibitem [{\citenamefont {Messiah}(2014)}]{messiah2014quantum}%
  \BibitemOpen
  \bibfield  {author} {\bibinfo {author} {\bibfnamefont {A.}~\bibnamefont
  {Messiah}},\ }\href@noop {} {\emph {\bibinfo {title} {Quantum mechanics}}}\
  (\bibinfo  {publisher} {Courier Corporation},\ \bibinfo {year}
  {2014})\BibitemShut {NoStop}%
\bibitem [{\citenamefont {Leclerc}\ \emph {et~al.}(2012)\citenamefont
  {Leclerc}, \citenamefont {Viennot},\ and\ \citenamefont
  {Jolicard}}]{leclerc2012role}%
  \BibitemOpen
  \bibfield  {author} {\bibinfo {author} {\bibfnamefont {A.}~\bibnamefont
  {Leclerc}}, \bibinfo {author} {\bibfnamefont {D.}~\bibnamefont {Viennot}}, \
  and\ \bibinfo {author} {\bibfnamefont {G.}~\bibnamefont {Jolicard}},\
  }\href@noop {} {\bibfield  {journal} {\bibinfo  {journal} {Journal of Physics
  A: Mathematical and Theoretical}\ }\textbf {\bibinfo {volume} {45}},\
  \bibinfo {pages} {415201} (\bibinfo {year} {2012})}\BibitemShut {NoStop}%
\bibitem [{\citenamefont {Lou}\ \emph {et~al.}(2025)\citenamefont {Lou},
  \citenamefont {Cheng}, \citenamefont {Zhao},\ and\ \citenamefont
  {Tong}}]{lou2025nonadiabatic}%
  \BibitemOpen
  \bibfield  {author} {\bibinfo {author} {\bibfnamefont {X.-Y.}\ \bibnamefont
  {Lou}}, \bibinfo {author} {\bibfnamefont {Y.-S.}\ \bibnamefont {Cheng}},
  \bibinfo {author} {\bibfnamefont {P.}~\bibnamefont {Zhao}}, \ and\ \bibinfo
  {author} {\bibfnamefont {D.}~\bibnamefont {Tong}},\ }\href@noop {} {\bibfield
   {journal} {\bibinfo  {journal} {Physical Review A}\ }\textbf {\bibinfo
  {volume} {112}},\ \bibinfo {pages} {022222} (\bibinfo {year}
  {2025})}\BibitemShut {NoStop}%
\bibitem [{\citenamefont {Gronwall}(1919)}]{gronwall1919note}%
  \BibitemOpen
  \bibfield  {author} {\bibinfo {author} {\bibfnamefont {T.~H.}\ \bibnamefont
  {Gronwall}},\ }\href@noop {} {\bibfield  {journal} {\bibinfo  {journal}
  {Annals of Mathematics}\ }\textbf {\bibinfo {volume} {20}},\ \bibinfo {pages}
  {292} (\bibinfo {year} {1919})}\BibitemShut {NoStop}%
\bibitem [{\citenamefont {Bellman}(1943)}]{bellman1943stability}%
  \BibitemOpen
  \bibfield  {author} {\bibinfo {author} {\bibfnamefont {R.}~\bibnamefont
  {Bellman}},\ }\href@noop {} {\bibfield  {journal} {\bibinfo  {journal} {Duke
  Math. J.}\ }\textbf {\bibinfo {volume} {10}},\ \bibinfo {pages} {643}
  (\bibinfo {year} {1943})}\BibitemShut {NoStop}%
\end{thebibliography}

%

\end{document}